\documentclass[prb,twocolumn, superscriptaddress, longbibliography]{revtex4-2}

\usepackage{
float, 
amsmath, 
amssymb, 
bbm, 
mathrsfs, 
bm, 
braket, 
color, 
graphicx, 
comment, 
amsfonts, 
dsfont, 
mathrsfs,
xfrac, 
}

\usepackage{color}
\usepackage[dvipsnames]{xcolor}

\usepackage[colorlinks, citecolor=blue, urlcolor=blue]{hyperref}

\usepackage[mathscr]{euscript}

\usepackage[normalem]{ulem}
\usepackage{graphicx}
\usepackage{subfigure}
\usepackage{float}

\DeclareMathAlphabet\mathbfcal{OMS}{cmsy}{b}{n}

\definecolor{cobalt}{rgb}{0.0, 0.28, 0.67}

\begin{document}

\title{Vortical currents and reciprocal relations for transport coefficients in the electron hydrodynamic regime}

\author{Nisarg Chadha}
\affiliation{Undergraduate Programme, Indian Institute of Science, Bangalore 560012, India}
\affiliation{Department of Physics, Indian Institute of Science, Bangalore - 560012, India}
\author{Subroto Mukerjee}
\affiliation{Department of Physics, Indian Institute of Science, Bangalore - 560012, India}
\date{\today}
\begin{abstract}
    We investigate the hydrodynamic regime in metals with momentum-conserving electron-electron scattering. 
    The conservation of momentum results in well-defined dynamics whose effects we investigate via the relevant continuity equations. 
    We find anomalous contributions to the charge and heat transport currents arising from gradients of the velocity field in a semiclassical treatment with a Berry curvature.
    These contributions are non-vanishing for systems lacking inversion symmetry, and the corresponding transport coefficients do not obey the standard Onsager reciprocity relations.
    Instead, we show that the response coefficients relating the currents to the stress tensor obey independent reciprocity relations with the stress tensor and thus exhibit cross-tensor effects of charge and heat transport with the momentum transport.
    The Berry curvature contribution to the stress magnetization tensor is also derived.
    
\end{abstract}
\maketitle

\textit{Introduction.} 
Electron transport in most metals is well understood to arise from diffusive scattering due to impurities or phonons.
However, the synthesis of high-purity 2D materials has now made it possible to probe the \textit{hydrodynamic regime} of electron dynamics postulated to exist by Gurzhi\cite{gurzhi1963minimum,gurzhi1968hydrodynamic} in early work.
The hydrodynamic regime is characterized by the domination of electron-electron interactions over impurity and phonon scattering, giving rise to a viscous electronic fluid with exotic collective dynamics.
This regime has striking novel transport signatures, some of which have been investigated theoretically\cite{PhysRevLett.106.256804,PhysRevLett.115.056603,PhysRevLett.118.226601, doi:10.1073/pnas.1612181114} and experimentally\cite{PhysRevB.49.5038,PhysRevB.51.13389,doi:10.1126/science.aac8385,doi:10.1126/science.aad0343, doi:10.1126/science.aad0201,krishna2017superballistic,sulpizio2019visualizing, aharon2022direct,kumar2022imaging}.

We shall be interested in the response to perturbations which vary slowly compared to the microscopic relaxation scales.
In this regime, only the conserved variables exhibit well-defined dynamics in the long-time limit\cite{chaikin1995principles}.
In conventional metals, the diffusive dynamics of charge and entropy densities are characterized by the charge and heat currents arising in response to driving forces.
These forces can be gradients in mechanical fields(like the electrical potential) controlling microscopic dynamics or statistical fields(temperature, chemical potential) appearing from the distribution function.
This hydrodynamic theory can be applied to a vast assortment of systems ranging from spin chains\cite{chaikin1995principles}, plasma magnetohydrodynamics\cite{goossens2003introduction}, and Bose-Einstein condensates\cite{tsubota2013quantum}, in addition to electron hydrodynamic systems.

The transport coefficients associated with the charge and heat currents obey thermodynamic relations such as the Einstein relations\cite{einstein1905motion,einstein1956investigations} and the Onsager reciprocity relations\cite{PhysRev.37.405,PhysRev.38.2265}.
Transport due to a mechanical field coupling to some variable is related to the transport due to the statistical field conjugate to the variable via the Einstein relations.
Onsager's relations\cite{PhysRev.37.405,PhysRev.38.2265,RevModPhys.17.343} relate cross-phenomena, such as charge currents due to temperature gradients(Seebeck effect) and heat currents due to potential gradients(Peltier effect).

Ref.~\cite{PhysRevB.55.2344} highlighted that these thermodynamic relations are valid only for the physically meaningful \textit{transport currents}, which are obtained by subtracting the \textit{magnetization currents} from the total currents.
Experimentally measured currents thus correspond to transport currents, which vanish in equilibrium.
The magnetization currents are associated with internal circulating currents, which do not correspond to any net transport.

The thermodynamic relations have also been shown\cite{PhysRevLett.97.026603} to hold for transport arising from geometric contributions\cite{RevModPhys.82.1959} to electron dynamics\cite{PhysRevB.53.7010, PhysRevB.59.14915}.
The quantum anomalous Hall effect\cite{PhysRevLett.61.2015},
spin Hall effect\cite{PhysRevLett.95.226801, PhysRevLett.97.026603}, valley Hall effect\cite{valleyhall}, and the anomalous Nernst effect\cite{PhysRevLett.115.246601} are examples of transport signatures arising from a Berry curvature.
Geometric effects on hydrodynamic transport have been shown\cite{PhysRevResearch.2.032021, PhysRevB.102.205201,PhysRevB.103.125106} to give an asymmetric Poiseulle flow profile and an anomalous charge vortical current arising from velocity gradients in the viscous fluid.

Given the theoretical and experimental interest surrounding hydrodynamics, there is a surprising gap in the discussion of certain important aspects of the framework of hydrodynamic transport in the existing literature.
A complete understanding of the thermodynamics of hydrodynamic systems is necessary to understand their novel transport signatures, such as the vortical current, whose thermodynamics cannot be studied within the framework developed so far.
Since momentum is conserved in the bulk, its dynamics must be included in the thermodynamic framework, leading to additional reciprocal relations between the transport coefficients.

In this Letter, we present a complete framework for transport in systems with number, energy, and momentum conservation. 
Our main results are:
1) A derivation of the {\em complete} set of reciprocity relations for charge, heat, and momentum transport, 2) an expression for the electronic stress magnetization, and 3) a demonstration of the existence of a heat vortical current analogous to the charge vortical current derived previously. 
Along the way, we formulate a theory for the expressions of current operators in the presence of external fields to derive the magnetization response.
We show that the vortical transport currents in the hydrodynamic system with a Berry curvature provide a non-trivial realization of the complete set of reciprocity relations.  

\textit{Reciprocal relations.} Onsager's reciprocity relations\cite{PhysRev.37.405, PhysRev.38.2265} relate cross-transport coefficients using the time-reversibility of the microscopic equations of motion.

Since momentum is a well-defined variable for hydrodynamic systems, its dynamics can be studied using the momentum current, given by the stress tensor.
In addition, the conjugate field to the momentum density- the velocity field- is a well-defined statistical field.
This implies that velocity gradients can drive currents, just like chemical potential and temperature gradients.
Although directly creating velocity gradients in electronic systems is complicated, they can naturally arise in the presence of external fields with an appropriate choice of boundary conditions\cite{PhysRevB.103.125106}.
Including these \textit{shear forces}, we can write the most general phenomenological expressions for all the currents in terms of all possible driving forces(gradients in the electric potential $\phi$, temperature $T$, and velocity $\mathbf{u}$) as\cite{PhysRev.94.218}:
\begin{subequations}\label{eq:phenomeno}
\begin{equation}
    j^N_i=-L^1_{i,j} \nabla_j \phi-L^2_{i,j}\nabla_j T-L^3_{i,jk}\partial_k u_j
\end{equation}
\begin{equation}
    j^Q_i=-L^4_{i,j}\nabla_j \phi-L^5_{i,j}\nabla_j T-L^6_{i,jk}\partial_k u_j
\end{equation}
\begin{equation}
    \Pi_{i,j}=-L^7_{ij,k}\nabla_k \phi-L^8_{ij,k}\nabla_k T-L^9_{ij,kl}\partial_l u_k
\end{equation}
\end{subequations}
$\mathbf{j^N}, \mathbf{j^Q}$ are the charge and heat currents respectively, and $\Pi$ is the second-rank stress tensor.
$i,j,k$ are the spatial indices($x,y,z$), and summation over repeated indices is assumed throughout this manuscript.
The transport coefficients with three indices describe vector currents due to tensor forces or vice-versa.
Thus, we call them($L^3,L^6,L^7$,$L^8$) cross-tensor coefficients.
Curie's symmetry principle\cite{de1951thermodynamics} states that such cross-tensorial coefficients vanish for an isotropic system.

The reciprocal relations are obtained using the microscopic time-reversibility of the equations of motion using the correlations between fluctuations in the conserved variables\cite{PhysRev.94.218}.
The presence of time-reversal symmetry breaking terms can be incorporated by accounting for the appropriate transformation of intrinsic as well as extrinsic parameters under time-reversal\cite{mazur1953onsager,PhysRevResearch.2.022009}.
The relevant quantities in the derivation are $n, s, \mathbf{g}$ the number, entropy and momentum density respectively and the explicit derivation is provided in the Supplemental Material. The resultant relations for the coefficients $L^3,L^6,L^7$,$L^8$ , which are the first main result of this work, are:
\begin{subequations}\label{eq:reciprocal}
    \begin{equation}
        L^3_{k,ij}=-L^7_{ij,k}
    \end{equation}
    \begin{equation}
        L^8_{ij,k}=-\frac{1}{T}L^6_{k,ij}
    \end{equation}
\end{subequations}

\textit{Transport and Magnetization Currents.} 
Collisions preserving energy, momentum, and charge density give rise to the corresponding conservation laws described by the continuity equations:
\begin{subequations}\label{eq:continuity}
\begin{equation}
    -e\partial_t \bar{n}+\nabla.\mathbf{J^N}=0
\end{equation}
\begin{equation}
    \partial_t \bar{g}_i+\nabla_j\Pi_{ij}=e\bar{n}\nabla_i\phi
\end{equation}
\begin{equation}
\partial_t \bar{\epsilon}+\nabla.\mathbf{J^E}=0
\end{equation}
\end{subequations}
Where $\bar{\epsilon}$ is the energy density, $\mathbf{J^E}$ is the energy current(related to the heat current as $\mathbf{J^Q}=\mathbf{J^E}+\frac{\mu}{e}\mathbf{J^N}$), and $\Pi_{ij}$ is the stress tensor.

It is well known that divergenceless magnetization currents need to be subtracted from the total currents obtained from the continuity equations to obtain transport currents~\cite{PhysRevB.55.2344}. It is the transport currents that obey the Onsager and Einstein relations\cite{einstein1905motion, einstein1956investigations} and vanish in equilibrium.

In equilibrium in the absence of external fields, the charge and energy magnetization currents are given by the respective curls of the magnetization densities $\mathbf{M}^N_0,\mathbf{M}^E_0$.
Analogously, we define the stress magnetization tensor $\mathbf{M}^{\pi}_0$ such that the magnetization stress tensor in equilibrium is given by $[\Pi_M]_{ij}=\epsilon_{jkl}\partial_k [M^{\pi}_0]_{il}$. The change in the magnetization under external fields must be accounted for to get the correct transport and magnetization currents.

For systems without momentum transport, the magnetization currents in the presence of external fields are obtained by subjecting the system to external electric and ``gravitational'' potentials~\cite{PhysRev.135.A1505,PhysRevB.55.2344}. The currents in the presence of chemical potential and temperature gradients can also be obtained this way via the Einstein relations. With momentum transport, the system also has to be subjected to a gradient in the {\em boost} potential ($\boldsymbol{\chi}$ conjugate to the momentum density as described in the supplemental material. The Hamiltonian density in the presence of external electric, gravitational, and a boost potentials is \cite{PhysRev.135.A1505}:

\begin{equation}\label{eq:ruzin}
\hat{h}_T(\mathbf{r})=\hat{h}(\mathbf{r})(1+\psi)-e\hat{n}(\mathbf{r})\phi+\hat{\mathbf{g}}(\mathbf{r}).\boldsymbol{\chi},
\end{equation}
where the number density $\hat{n}(\mathbf{r})=\Sigma_i \delta(\mathbf{r}-\mathbf{\hat{r}}_i)$, momentum density $\mathbf{\hat{g}}(\mathbf{r})=1/2\Sigma_i \{\mathbf{\hat{p}}_i,\delta(\mathbf{r}-\mathbf{\hat{r}}_i)\}$, and the energy density $\hat{h}(r)=1/2\Sigma_i\{\hat{h}_i,\delta(\mathbf{r}-\mathbf{\hat{r}}_i)\}$ and $\hat{h}_i$ is the single particle Hamiltonian, appropriately defined to include interactions(see the supplemental material).

The continuity equations Eqn.~\ref{eq:continuity} along with Eqn.~\ref{eq:ruzin} can be used to obtain expressions for the currents densities from which one can obtain expressions for the magnetization currents following the procedure outlined in Ref.~\cite{PhysRevB.55.2344} (see Supplemental Material). The expressions are   
\begin{subequations}\label{eq:magnetizationfluxes}
    \begin{equation}
    (\mathbf{J}^N_M)_i=[\mathbf{J}^N_M]_i+(\nabla\psi\times\mathbf[{M}^N_0])_i
    \end{equation}
    \begin{equation}
        (\mathbf{J}^E_M)_i=[\mathbf{J}^E_M]_i+2(\nabla\psi\times\mathbf[{M}^E_0])_i+(\nabla\phi\times[\mathbf{M}^N_0])_i+\epsilon_{ijl}\partial_j\chi_k[M^{\pi}_0]_{kl}
    \end{equation}
    \begin{equation}
        (\Pi_M)_{ij}=[\Pi_M]_{ij}+\epsilon_{jkl}\partial_k \psi [\mathbf{M}^{\pi}_0]_{il},
    \end{equation}
\end{subequations}
where $[ \dots ]$ denotes the equilibrium value of a particular quantity. For systems with momentum transport, the existence of the stress magnetization tensor $(M^{\pi}_0)_{ij}$ and the associated magnetization current $(\Pi_M)_{ij}$ is the second important result of this paper. Ref.~\cite{10.21468/SciPostPhysCore.6.3.052} derived the expressions for the magnetization densities by constraining the transport currents to obey the Einstein relations.
We shall use the same approach to derive an expression for $(M^{\pi}_0)_{ij}$.

\textit{Boltzmann transport formalism for hydrodynamic systems.} 
Having laid out the formalism to obtain magnetization currents, we shall now derive explicit expressions for the total currents of the system in terms of the external fields from which the magnetization currents will be subtracted to obtain transport currents. For this, we employ the semiclassical Boltzmann transport formalism for Bloch wave packets~\cite{girvin2019modern}.
The equations of motion obeyed by the centre of the wave packet in the presence of an electric field are:
    $\mathbf{\dot{p}_c}=-e\mathbf{E},
    \mathbf{\dot{r}_c}=\frac{\partial\epsilon_{\mathbf{p}}}{\partial \mathbf{p}}-\frac{1}{\hbar}\dot{\mathbf{p}}_c\times\mathbf{\Omega_{p}}$.
Here, $\epsilon_\mathbf{p}$ is the band dispersion responsible for the usual group velocity, $\Omega_{\mathbf{p}}$, the Berry curvature which gives rise to the anomalous velocity, and $(\mathbf{r_c},\mathbf{p_c})$ denotes the position of the wave packet centre in phase space.
Hereafter, we shall drop the subscripts on the wave packet coordinates. We also consider the dynamics in just a single band even though a minimum of two are required for a system to have a non-zero Berry curvature. We assume that the energy difference between the Fermi energy and the closest energy in any other band is sufficiently larger than the temperature and the inter-band scattering rate (converted to energy units) to ignore the presence of other bands.
\begin{figure*}[htp]
\centering
\includegraphics[width=\textwidth]{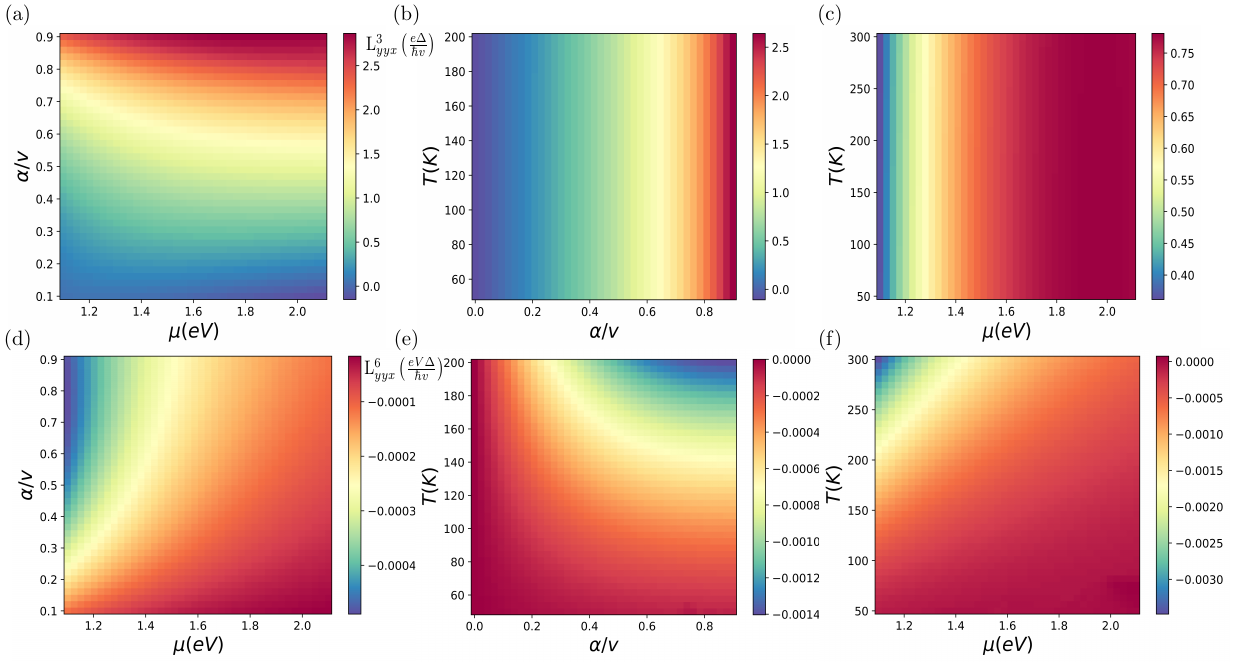}
\caption{$L^3_{yyx}$ (top) and $L^6_{yyx}$ (bottom): The figures show the variation of the cross-tensor coefficients for the Hamiltonian in Eq.~\eqref{eq:ham2d} as functions of the tilt $\alpha$, chemical potential $\mu$, and temperature $T$. We choose the parameters $\Delta=1eV$, and $v=10^5 ms^{-1}$. Figures (a) and (d) are for $T=100K$, (b) and (e) are for $\mu=1.3eV$, and (c) and (f) are for $\alpha=0.4v$. $L^3$ exhibits a very weak temperature dependence for the chosen range of temperatures($T\ll\mu, \Delta$). The scaling of the coefficients has been further discussed in the supplemental material.} 
\label{fig:crosstens}
\end{figure*}
Bulk momentum conservation considerably simplifies the form for the distribution function $f(\mathbf{r},\mathbf{p},t)$ due to a well-defined conjugate velocity field $\mathbf{u}(\mathbf{r})$.
Consider a small volume element of the system centred about $\mathbf{r}$ with an average velocity $\mathbf{u}(\mathbf{r})$.
The lab Hamiltonian in terms of the coordinates of the boosted fluid frame is $h(\mathbf{r})-\mathbf{u}.\mathbf{g}$, as in Eq.~\eqref{eq:ruzin}.
Since the system is in equilibrium in the fluid's local rest frame, we can use the distribution function\cite{chaikin1995principles, lucas2018hydrodynamics, PhysRevB.103.125106}:
\begin{equation}\label{eq:distribution}
    f(\mathbf{r},\mathbf{p},t)=\frac{1}{e^{\beta(\mathbf{r},t)(\epsilon_{\mathbf{p}}-\mu(\mathbf{r},t)-\mathbf{u}(\mathbf{r}).\mathbf{p})}+1}
\end{equation}
The velocity profile can be explicitly found by using the Navier-Stokes equation obtained by substituting Eq.~\eqref{eq:distribution} in Eq.~\eqref{eq:continuity}\cite{PhysRevB.103.125106}.
Eq.~\eqref{eq:distribution} allows us to compute the expressions for the total current densities, defined as:
\begin{equation}\label{eq:currents}
\mathcal{O}=\int{\frac{[d\mathbf{p}]}{\hbar^2}\chi\left(\frac{\partial\epsilon_{\mathbf{p}}}{\partial \mathbf{p}}+e\hbar\mathbf{E}\times\Omega_{\mathbf{p}}\right)f(\mathbf{r},\mathbf{p},t)}
\end{equation}
with $\mathcal{O}=\{\mathbf{J^N},\mathcal{J^Q},\Pi\}$ being the currents associated with the variables $\chi=\{-e,s,\mathbf{p}\}$.
Here $[d\mathbf{p}]=\frac{d^2p}{(2\pi)^2}$ is the integration measure over \textbf{p}-space.
Since the energy spectrum is defined up to a constant, it is the heat current rather than the energy current, which is physically relevant. 

The charge and energy magnetization are given by\cite{PhysRevLett.97.026603,PhysRevB.102.235161,10.21468/SciPostPhysCore.6.3.052}:
\begin{subequations} \label{eq:magnande}
    \begin{equation}
        \mathbf{M}^N_0=\frac{e}{\hbar}\int{[d\mathbf{p}] k_B T\Omega_{\mathbf{p}}\log(1+e^{-\beta(\epsilon_{\mathbf{p}}-\mu-\mathbf{u}.\mathbf{p})})}
    \end{equation}
    \begin{equation}
    \mathbf{M}^E_0=-\frac{1}{\hbar}\int{[d\mathbf{p}]\Omega_{\mathbf{p}}\int_{-\infty}^\mu d\tilde{\mu}[\epsilon_{\mathbf{p}}f(\tilde{\mu})+k_B T\log(1+e^{-\beta(\epsilon_{\mathbf{p}}-\tilde{\mu}-\mathbf{u}.\mathbf{p})})]}
    \end{equation}
\end{subequations}
Here, $f(\tilde{\mu})$ refers to the distribution function in Eq.~\eqref{eq:distribution} with $\mu=\tilde{\mu}$.
In the $\mathbf{u}=0$ limit, the expressions match the known expressions for equilibrium magnetization densities\cite{10.21468/SciPostPhysCore.6.3.052}.
Although we have taken $\psi=\boldsymbol{\chi}=0$ for obtaining the expressions for the stress magnetization and the transport currents, we have derived the Einstein relation for $\boldsymbol{\chi}$ and verified that our expressions satisfy the other two Einstein relations in the supplemental Material. The transport currents are obtained by subtracting the magnetization currents from the total currents in Eq.~\eqref{eq:currents} to linear order (see Supplemental material). We obtain the expression for the stress magnetization density by demanding that the transport fluxes only depend on the combination $\nabla\mu+e\mathbf{E}$, as dictated by the Einstein relation.
This gives the second main result of our work- the expression for the stress magnetization tensor(derived in the Supplemental Material). 
\begin{equation}\label{eq:magpi}
    \mathbf{M}_0^{\pi}=-\frac{1}{\hbar}\int{[d\mathbf{p}]\mathbf{p}\otimes\Omega_{\mathbf{p}}k_B T\log(1+e^{-\beta(\epsilon_{\mathbf{p}}-\mu-\mathbf{u}.\mathbf{p})})}
\end{equation} \label{eq:magpi}
Where $\otimes$ represents the dyadic tensor product of the two vectors($[\mathbf{p}\otimes\Omega_{\mathbf{p}}]_{ij}=p_i[\Omega_{\mathbf{p}}]_j$).
Using Eqns.~\ref{eq:magnande} and ~\ref{eq:magpi} along with the expressions for the transport currents in terms of the magnetization densities, we get the third main result of our work- the complete expressions for the charge and heat transport currents and transport stress tensor:
\begin{widetext}
\begin{subequations}\label{eq:transport_stress_heat}
\begin{equation}
 \begin{aligned}
        \mathbf{J}^N_T=-e\bar{n}\mathbf{u}-\frac{e^2}{\hbar}\mathbf{E}\times\int{[d\mathbf{p}]\Omega_{\mathbf{p}}f^0(\mathbf{r},\mathbf{p},t)-\frac{e}{\hbar}\nabla T\times\int{[d\mathbf{p}]}\Omega_{\mathbf{p}}\left[\frac{\epsilon_{\mathbf{p}}-\mu}{k_B T}f^0+\log(1+e^{-\beta(\epsilon_\mathbf{p}-\mu)})\right]}-\frac{e}{\hbar}\int{[d\mathbf{p}]\nabla(\mathbf{p}.\mathbf{u})\times\Omega_{\mathbf{p}}f^0}
    \end{aligned}
\end{equation}
\begin{equation}
    \begin{aligned}
        \mathbf{J}^Q_T=T\bar{s}\mathbf{u}+\frac{e^2}{\hbar}k_B T\mathbf{E}\times\int{[d\mathbf{p}]\left[\frac{\epsilon_{\mathbf{p}}-\mu}{k_B T} f^0+\log(1+e^{-\beta(\epsilon_\mathbf{p}-\mu)})\right]}+\frac{1}{\hbar}\frac{\nabla T}{T}\times\int{[d\mathbf{p}]\Omega_{\mathbf{p}}\int{d\tilde{\mu}(\epsilon_{\mathbf{p}}-\tilde{\mu})^2\frac{\partial f^0}{\partial \tilde{\mu}}}}+\\\frac{1}{\hbar}k_B T\int{[d\mathbf{p}]}\nabla(\mathbf{p.u})\times\Omega_{\mathbf{p}}\left[\frac{\epsilon_{\mathbf{p}}-\mu}{k_B T}f^0+\log(1+e^{-\beta(\epsilon_{\mathbf{p}}-\mu)})\right]
    \end{aligned}
    \end{equation}
\begin{equation}
    \begin{aligned}
        [\Pi_T]_{ij}=P\delta_{ij}+\frac{e}{\hbar}\epsilon_{jkl}\int{[d\mathbf{p}]p_iE_k[\Omega_{\mathbf{p}}]_lf^0}+\frac{1}{\hbar}\epsilon_{jkl}\frac{\partial_k T}{T}\int{[d\mathbf{p}]p_i[(\epsilon_{\mathbf{p}}-\mu)f^0+k_BT\log(1+e^{-\beta(\epsilon_{\mathbf{p}}-\mu)})][\Omega_{\mathbf{p}}]_l},
    \end{aligned}
\end{equation}
\end{subequations}
\end{widetext}
where $P=k_B T\int{[d\mathbf{p}]\log(1+e^{-\beta(\epsilon_{\mathbf{p}}-\mu)})}$ is the pressure. 

Along with the usual drift currents, the Berry curvature generates transverse currents responsible for the anomalous Hall effect, anomalous Nernst effect, anomalous Seebeck effect, and anomalous thermal Hall effect.
Additionally, we obtain current contributions that involve a coupling between the Berry curvature and velocity gradients.
The charge current in Eq.~\eqref{eq:transport_stress_heat}a has been discussed in Ref.~\cite{PhysRevResearch.2.032021} and has analogues in the context of vortical currents in chiral rotating relativistic fluids\cite{PhysRevLett.103.191601}.
Here, we also derive the thermal analog of the vortical current.
We can extract the expressions for the cross-tensor coefficients defined in Eq.~\eqref{eq:phenomeno} from Eqn.~\eqref{eq:transport_stress_heat},

\begin{subequations}\label{eq:vortcoeffs}
    \begin{equation}
        L^3_{i,jk}=\frac{e}{\hbar}\epsilon_{ikl}\int{[d\mathbf{p}]p_j\Omega_{\mathbf{p},l}f^0}
    \end{equation}
    \begin{equation}
    \begin{split}
        L^6_{i,jk}=-\frac{k_B T}{\hbar}\epsilon_{ikl}\int[d\mathbf{p}]p_j\Omega_{\mathbf{p},l}\left[\frac{\epsilon_{\mathbf{p}}-\mu}{k_B T}f^0+\right.\\\left.\log(1+e^{-\beta(\epsilon_{\mathbf{p}}-\mu)})\vphantom{\int_1^2}\right].
    \end{split}
    \end{equation}
    \begin{equation}
        L^7_{ij,k}=\frac{e}{\hbar}\epsilon_{jkl}\int{[d\mathbf{p}]p_i\Omega_{\mathbf{p},l}f^0}
    \end{equation}
    \begin{equation}
    \begin{split}
        L^8_{i,jk}=-\frac{k_B}{\hbar} \epsilon_{jkl}\int[d\mathbf{p}]p_i\Omega_{\mathbf{p},l}\left[\frac{\epsilon_{\mathbf{p}}-\mu}{k_B T}f^0+\right.\\\left.\log(1+e^{-\beta(\epsilon_{\mathbf{p}}-\mu)})\vphantom{\int_1^2}\right]
    \end{split}
    \end{equation}
\end{subequations}
Thus, the transport coefficients associated with the charge and heat vortex currents obey the cross-tensor reciprocity relations Eq.~\eqref{eq:reciprocal}.
The contribution to the transport stress tensor from the electric field comes through the anomalous velocity, whereas the contribution from the temperature gradient appears by subtracting the magnetization stress tensor.
Note that for a 2D system in a bar geometry, the shear force is perpendicular to the applied field and hence the vortex currents will be longitudinal relative to the applied field.
Moreover, for an isotropic system, inversion symmetry dictates $\Omega_{\mathbf{p}}=\Omega_{\mathbf{-p}}$, which makes the cross-tensor coefficients vanish. 
This is a manifestation of Curie's symmetry principle for the anomalous transport caused by the velocity gradients.

\textit{Calculation for a microscopic model:} Now we calculate the coefficients for a microscopic model in two dimensions given by Ref.\cite{PhysRevLett.115.216806}.
\begin{equation} \label{eq:ham2d}
     H = vp_x\sigma_y - sp_y \left(v\sigma_x - \alpha \right) + \Delta \sigma_z,
\end{equation}

$\Delta$ introduces a gap in the spectrum for the two valleys labeled by $\tau_z$, $v$ is the dispersion velocity in the absence of a gap and $\alpha$ introduces a tilt in the Dirac cones. $s=\pm 1$ marks the valley index. 

The vector ${\bf p=\textcolor{cyan}{\hbar}\bf k}$ is measured about the Dirac points for each of the two valleys.

The energy dispersion for the two valleys is
\begin{equation} \label{eq:disp}
    \epsilon_s({\bf p}) = s \alpha p_y + {\rm sgn} (\mu) \sqrt{\left(v^2p^2 + \Delta^2\right)},
\end{equation}
where the chemical potential $\mu > 0 (< 0)$ corresponds to the conduction (valence) band. The Berry curvature
\begin{equation} \label{eq:Berry}
    \Omega_s ({\bf p}) = \frac{{\rm sgn}(\mu)} {2} \frac{sv^2\Delta}{\left(v^2p^2 + \Delta^2\right)^{3/2}}
\end{equation}
Considerations of symmetry imply that only $L^3_{y,yx}=-L^3_{x,yy}$ and $L^6_{y,yx}=-L^6_{x,yy}$ are non-zero. 
Fig. \ref{fig:crosstens} shows the results for the numerical computation of $L^3_{y,yx}$ and $L^6_{y,yx}$. 

In the limit $\alpha\ll v, \Delta/k_F$, we can obtain low temperature behavior for the vortical coefficients(see Supplemental material for details). 
At T=0, $L^3_{yyx}$ is given by:
\begin{equation} \label{eq:vort1}
L^3_{x,yy} = \frac{e}{4\pi\hbar} \frac{ \Delta \alpha (\mu^2-\Delta^2)}{v^2\mu^2},
\end{equation}

Since thermal transport vanishes at $T=0$, we perform a Sommerfeld expansion to get the leading order temperature dependence for $L^6$ in the limit $\alpha \ll v, \Delta/p_F$(see Supplemental material for derivation):
\begin{equation}\label{eq:l6approx}
    L^6_{xyy}=\frac{\pi\alpha \Delta^3}{6\hbar\mu^3v^2}(k_B T)^2
\end{equation}

\textit{Experimental realization:} 
Although visualizing momentum flow in electronic systems remains an open problem, detecting the vortical currents in charge and heat transport provides an avenue to study anomalous cross-tensor transport.
The most natural setting to obtain these currents would be to setup a Couette flow in the electronic system.
However, the constraint by the underlying solid lattice makes it difficult to realize the required boundary conditions.
Instead, we can indirectly realize velocity gradients through the intrinsic electronic viscosity in the presence of a driving field.
As discussed for the charge vortical current\cite{PhysRevResearch.2.032021}, bilayer graphene and $\mathrm{WP_2}$ are experimental candidates to realize anomalous hydrodynamics in non-centrosymmetric systems.
Although we have used the Berry curvature to drive currents via the anomalous velocity, shear forces in the presence of time-reversal breaking through a magnetic field, angular velocity, or spin-orbit coupling can also drive vortex currents.

\textit{Summary.} 
In summary, we have developed a complete framework to describe transport in momentum-conserving systems.
We identify the central role played by the velocity gradients as shear forces to drive currents in the system.
Although our work is motivated by electron hydrodynamic systems, the cross-tensor reciprocity relations and expressions for magnetization densities are general for momentum-conserving systems.
For instance, the vortical currents we have discussed have counterparts in rotating, relativistic, chiral fluids\cite{PhysRevLett.103.191601}.

By applying our formalism to an electronic system with a non-trivial Berry curvature, we have obtained a heat vortical current alongside the known charge vortical current, arising from shear forces coupling to the Berry curvature.
The vortical currents obey the cross-tensor reciprocity relations with the transport stress tensor.
We encourage experimental imaging of anomalous cross-tensor charge and heat transport through imbalance valleytronic/spintronic measurements.

\textit{Acknowledgements}
NC acknowledges support from the Kishore Vaigyanik Protsahan Yojana from the Dept. of Science and Technology (DST), Govt. of India. SM thanks the DST, Govt. of India for support.

\bibliography{references}

\end{document}


\title{Supplemental Material}
\author{Nisarg Chadha}
\affiliation{Undergraduate Programme, Indian Institute of Science, Bangalore 560012, India}
\affiliation{Department of Physics, Indian Institute of Science, Bangalore - 560012, India}
\author{Subroto Mukerjee}
\affiliation{Department of Physics, Indian Institute of Science, Bangalore - 560012, India}
\date{\today}
\maketitle

\section{Derivation of the Reciprocal Relations}
We derive the reciprocal relations for an adiabatic system perturbed from equilibrium, adopting the notation from Ref.~\cite{PhysRev.94.218}.
The state variables can be divided into those which are even under time-reversal($A_i$) and those which are odd($B_i$).
Define the deviation of the respective variables from their equilibrium values $A_i^0$ and $B_i^0$ as $\alpha_i$ and $\beta_i$ respectively.
To lowest order, the change in entropy due to the perturbation is of the form(assuming double summation over repeated indices):
\begin{equation}
    \Delta S=-\frac{1}{2}\int{dV dV'(g_{ij}(\mathbf{r},\mathbf{r'})\alpha_i(\mathbf{r})\alpha_j(\mathbf{r'})+h_{kl}(\mathbf{r},\mathbf{r
    })\beta_k(\mathbf{r})\beta_l(\mathbf{r'}))}
\end{equation}
The negative sign is a consequence of the second law of thermodynamics, since $\Delta S\leq 0$($g(\mathbf{r},\mathbf{r'})$ and $h(\mathbf{r},\mathbf{r'})$ are positive semi-definite).
The distribution for the fluctuations is given by:
\begin{equation}\label{eq:prob}
    P(\{\alpha,\beta\})=\frac{e^{\Delta S/k_B}}{\int{\Pi_{i,j} d\alpha_i d\beta_j e^{\Delta S/k_B}}}
\end{equation}

Defining the conjugate variables as respective functional derivatives:
\begin{align}\label{eq:conj_avg}
    X_i(\mathbf{r})=\frac{\partial\Delta S}{\partial\alpha_i}=-\int{dV' g_{ij}(\mathbf{r},\mathbf{r
    '})\alpha_j(\mathbf{r'})} &&
    Y_k(\mathbf{r})=\frac{\partial\Delta S}{\partial\beta_k}=-\int{dV' h_{kl}(\mathbf{r},\mathbf{r'
    })\beta_l(\mathbf{r'})}
\end{align}
Using Eq.~\eqref{eq:prob}, we get the equal-time thermodynamic averages:
\begin{align}\label{eq:conj_avg}
\langle\alpha_i(\mathbf{r})X_j(\mathbf{r'})\rangle=-k_B\delta_{ij}\delta(\mathbf{r}-\mathbf{r'}) &&
\langle\beta_k(\mathbf{r})Y_l(\mathbf{r'})\rangle=-k_B\delta_{kl}\delta(\mathbf{r}-\mathbf{r'})
\end{align}
Other combinations of state variables and conjugate variables vanish since the averages between $\alpha$ and $\beta$ variables vanish.
Eq.~\eqref{eq:conj_avg} implies\cite{PhysRev.94.218}:
\begin{align}\label{eq:conj_avg2}
\langle\alpha_i(\mathbf{r})\mathcal{D}(\mathbf{r'})X_j(\mathbf{r'})\rangle=-k_B\delta_{ij}\mathcal{D}(\mathbf{r'})\delta(\mathbf{r}-\mathbf{r'}) &&
\langle\beta_k(\mathbf{r})\mathcal{D}(\mathbf{r'})Y_l(\mathbf{r'})\rangle=-k_B\delta_{kl}\mathcal{D}(\mathbf{r'})\delta(\mathbf{r}-\mathbf{r'})
\end{align}
Where $\mathcal{D}(\mathbf{r})$ is of the form $\Sigma a_{ijk}\partial_i\partial_j\partial_k$.

Onsager used the time-reversibility of the equations of motion to get the relations\cite{PhysRev.37.405,PhysRev.94.218}:
\begin{subequations}\label{eq:onsager}
\begin{equation}
    \langle\alpha_i(\mathbf{r},t)\partial_t\alpha_j(\mathbf{r'
    },t)\rangle=\langle\alpha_j(\mathbf{r'},t)\partial_t\alpha_i(\mathbf{r},t)\rangle
\end{equation}
\begin{equation}
    \langle\beta_k(\mathbf{r},t)\partial_t\beta_l(\mathbf{r'
    },t)\rangle=\langle\beta_l(\mathbf{r'},t)\partial_t\beta_k(\mathbf{r},t)\rangle
\end{equation}
\begin{equation}
    \langle\alpha_i(\mathbf{r},t)\partial_t\beta_k(\mathbf{r'
    },t)\rangle=-\langle\beta_k(\mathbf{r'},t)\partial_t\alpha_i(\mathbf{r},t)\rangle
\end{equation}
\end{subequations}
The negative sign in the last equation appears due to opposite transformations of $\alpha$ and $\beta$ under time-reversal.

For the hydrodynamic system, the charge density $\rho=-en$ and entropy density $s$ are the $\alpha$ variables, and the components of the momentum density $\mathbf{g}$ are the $\beta$ variables.
The corresponding conjugate variables will be the chemical potential $\mu$ and $T$ as the $X$ variables, and the components of the velocity field $\mathbf{u}$ as the $Y$ variables.
Taking different pairs of variables in Eq.~\eqref{eq:onsager}, we can get the reciprocity relations.
\begin{equation}
    \langle\Delta \rho(\mathbf{r})\partial_t \Delta g_i(\mathbf{r'})\rangle=-\langle\Delta g_i(\mathbf{r'})\partial_t\Delta \rho(\mathbf{r'})\rangle
\end{equation}
Substituting the partial derivatives with divergences of currents using Eq.4 in the main text,
\begin{equation}
    \langle\Delta \rho(\mathbf{r})\partial_j'\Pi_{ij}(\mathbf{r'})\rangle=-\langle\Delta g_i(\mathbf{r'})\partial_j J^N_j(\mathbf{r'})\rangle
\end{equation}
$\partial'$ denotes derivatives over the primed coordinates.
Using the phenomenological relations in Eq.1 in the main text and Eq.~\eqref{eq:conj_avg}, we get:
\begin{equation}
    \langle\Delta\rho(\mathbf{r})\partial_j' L^7_{ij,k}(\mathbf{r'})\partial_k'\phi(\mathbf{r'})\rangle=-\langle \Delta g_i(\mathbf{r'})\partial_j L^3_{j,lk} \partial_k u_l(\mathbf{r})\rangle
\end{equation}
Using Eq.~\eqref{eq:conj_avg2},
\begin{equation}
    \partial_j' L^7_{ij,k}(\mathbf{r'})\partial_k'\delta(\mathbf{r}-\mathbf{r'})=-\partial_j L^3_{j,ik}\partial_k \delta(\mathbf{r}-\mathbf{r'})
\end{equation}
This implies, where $f$ is an analytic function, and the integral is over all space.
\begin{equation}
    \int{d^d\mathbf{r'} f(\mathbf{r'})\partial_j' L^7_{ij,k}(\mathbf{r'})\partial_k'\delta(\mathbf{r}-\mathbf{r'})}=-\int{d^d\mathbf{r'} f(\mathbf{r'})\partial_j L^3_{j,ik}\partial_k \delta(\mathbf{r}-\mathbf{r'})}
\end{equation}
Integrating by parts twice on the left(assuming that the transport coefficients vanish outside the sample volume), 
\begin{equation}
    \partial_j L^3_{j,ik}\partial_k f(\mathbf{r})=-\partial_k L^7_{ij,k}\partial_j f(\mathbf{r})
\end{equation}
Since $f$ is arbitrary here, we must have:
\begin{equation}
    L^3_{k,ij}=-L^7_{ij,k}
\end{equation}
Repeating the analysis for $\Delta s$ instead of $\Delta n$, we get:
\begin{equation}
    \frac{1}{T}L^6_{k,ij}=-L^8_{ij,k}
\end{equation}

\section{Current operators in the presence of external fields}
We derive the expressions of currents to first-order in external fields.
Although the formulation has been previously discussed\cite{PhysRev.135.A1505,PhysRevB.55.2344}, we present it again with the addition of the vector potential to discuss hydrodynamic systems.
Since we are interested in anomalous transport, we shall assume the absence of a magnetic field, although it can be easily incorporated into the formalism.
Consider a Hamiltonian of the form:
\begin{equation}
    H_0=\Sigma_i \hat{h}_i=\Sigma_i \left(\frac{\hat{p}_i^2}{2m}+V(\hat{\mathbf{r}})+\frac{1}{2}\Sigma_{j\neq i} u_{ij}\right)
\end{equation}
Where $V$ is the potential in the equilibrium system, and $u$ is the inter-particle interaction.
The operator forms for the state variables are given by:
\begin{align}\label{eq:dens_op}
    \hat{\rho}(\mathbf{r})=-e\Sigma_i \delta(\mathbf{r}-\mathbf{\hat{r}_i}) && \hat{h}(\mathbf{r})=\frac{1}{2}\Sigma_i \{\hat{h_i},\delta(\mathbf{r}-\mathbf{\hat{r}_i})\} && \hat{g}(\mathbf{r})=\frac{1}{2}\Sigma_i \{\hat{p_i},\delta(\mathbf{r}-\mathbf{\hat{r}_i})\}
\end{align}
External fields that couple to the charge and energy density are the electric potential $\phi$ and gravitational potential $\psi$ respectively\cite{PhysRev.135.A1505}.
Analogously, we introduce a vector potential($\boldsymbol{\chi}$) coupling to the momentum density.

In the presence of $\phi$, $\psi$, $\boldsymbol{\chi}$, the total Hamiltonian becomes $H_T=\int{d\mathbf{r}h_T(\mathbf{r})}$:
\begin{equation}
h_T(\mathbf{r})=h(\mathbf{r})(1+\psi)+\rho(\mathbf{r})\phi(\mathbf{r})+\mathbf{g}(\mathbf{r}).\boldsymbol{\chi}(\mathbf{r})
\end{equation}
Using Heisenberg's equation of motion, and the continuity equations, we can obtain the expressions for the current operators in the presence of external fields\cite{PhysRevB.55.2344}:
\begin{equation}
    \frac{d\rho}{dt}=-\frac{i}{\hbar}[\rho,H_T]=-\nabla.\mathbf{J^N}
\end{equation}
And similarly for the energy and momentum density.
Assuming the external fields to be uniform, we get expressions for the currents in terms of the currents in the absence of the external fields.
We take the equilibrium averages to get the relations:
\begin{subequations}\label{eq:eq_mag}
    \begin{equation}
        \langle\mathbf{J^N}\rangle=\langle\mathbf{j^N}\rangle(1+\psi)+\bar{\rho}\boldsymbol{\chi}
    \end{equation}
    \begin{equation}
        \langle J^E_i\rangle=\langle j^E_i\rangle(1+2\psi)+\langle j^N_i\rangle\phi+\langle\pi_{ji}\rangle \chi_j+\bar{\epsilon}\chi_i
    \end{equation}
    \begin{equation}
        \langle\Pi_{ij}\rangle=\langle\pi_{ij}\rangle(1+\psi)+\bar{g_i} \chi_j
    \end{equation}
\end{subequations}
Note that a non-zero $\boldsymbol{\chi}$ gives a transport current equal to $\bar{\mathcal{O}}\boldsymbol{\chi}$, where $\mathcal{O}$ is the variable whose current we are calculating and the bar denotes equilibrium averages.
The other terms involving combinations of currents give the corrections to the magnetization currents since the transport currents in the unperturbed system vanish\cite{PhysRevB.55.2344}.
The magnetization currents in the absence of the fields can be written as:
\begin{subequations}
    \begin{equation}
        [\mathbf{j^N_M}]_i=(\nabla\times\mathbf{M^N})_i=\partial_j A_{ij}^N; A^N_{ij}=-A^N_{ji}
    \end{equation}
     \begin{equation}
        [\mathbf{j^E_M}]_i=(\nabla\times\mathbf{M^E})_i=\partial_j A_{ij}^E; A^E_{ij}=-A^E_{ji}
    \end{equation}
     \begin{equation}
        [\mathbf{\Pi_M}]_{ij}=(\nabla\times\mathbf{M^{\pi}_i})_j=\partial_k A_{ijk}^{\pi}; A^{\pi}_{ijk}=-A^{\pi}_{ikj}
    \end{equation}
\end{subequations}
We use the $A$ tensors to make the notation convenient for the stress tensor.

For the $A$ tensors, Eq.~\eqref{eq:eq_mag} implies:
\begin{subequations}
    \begin{equation}
        \tilde{A}^N_{ij}=(1+\psi)A^N_{ij}
    \end{equation}
    \begin{equation}
        \tilde{A}^E_{ij}=(1+2\psi)A^E_{ij}+\phi A^N_{ij}+A^{\pi}_{kij}\chi_k
    \end{equation}
    \begin{equation}
        \tilde{A}^{\pi}_{ijk}=(1+\psi)A_{ijk}
    \end{equation}
\end{subequations}
For non-uniform fields that vary sufficiently slowly, Eq.~\eqref{eq:eq_mag} remains true to first order in the fields\cite{PhysRevB.55.2344}.
The magnetization currents in non-uniform fields, therefore become:
\begin{empheq}[box=\fbox]{gather}\label{eq:magcurr}
\mathbf{J^N_M}=[\mathbf{j^N_M}]+\nabla\psi\times[\mathbf{M^N_0}]\\
[J^E_M]_i=[j^E_M]_i+2\epsilon_{ijk}\partial_j\psi[M^E_0]_k+\epsilon_{ijk}\partial_j\phi [M^N_0]_k+\epsilon_{ijl}\partial_j \chi_k [M^{\pi}_0]_{kl}\\
[\Pi_M]_{ij}=[\pi_M]_{ij}+\epsilon_{jkl}\partial_k\psi [M^{\pi}]_{il}
\end{empheq}

\section{Stress Tensor Magnetization}
Using Eq.~\eqref{eq:magcurr}, we can obtain the transport currents by subtracting the magnetization currents from the total current.
Whereas total currents obtained from a coarse-grained analysis of the electron density in a wave packet acquire contributions from the orbital magnetic moment, these get cancelled from corresponding terms in the magnetization current\cite{PhysRevLett.97.026603}.
To keep the discussion centred on the magnetization contribution from the Berry curvature, we shall ignore these terms by defining the total currents as $\mathcal{F}^{\mathcal{O}}=\int{[d\mathbf{p}]f(\mathbf{r},\mathbf{p},t)\mathcal{O}(\frac{\partial\epsilon_{\mathbf{p}}}{\partial\mathbf{p}}+\frac{e}{\hbar}\mathbf{E}\times\Omega_{\mathbf{p}})}$, where $\mathcal{O}=\rho, \epsilon_{\mathbf{p}}$, and $\mathbf{p}$, and the respective currents $\mathcal{F}$ are $\mathbf{J^N}, \mathbf{J^E}$, and $\Pi$.
The expressions for the transport currents obtained after evaluating $\mathcal{F}^{\mathcal{O}}$ and subtracting the magnetization currents are

\begin{subequations}\label{eq:transportcurrents}
    \begin{equation}
        \mathbf{J}^N_T=-e\bar{n}\mathbf{u}-\frac{e^2}{\hbar}\mathbf{E}\times\int{[d\mathbf{p}]\Omega_{\mathbf{p}}f^0(\mathbf{r},\mathbf{p},t)-\frac{e}{\hbar}\nabla T\times\int{[d\mathbf{p}]}\Omega_{\mathbf{p}}\left[\frac{\epsilon_{\mathbf{p}}-\mu}{k_B T}f^0+\log(1+e^{-\beta(\epsilon_\mathbf{p}-\mu)})\right]}-\frac{e}{\hbar}\int{[d\mathbf{p}]\nabla(\mathbf{p}.\mathbf{u})\times\Omega_{\mathbf{p}}f^0}
    \end{equation}
    \begin{equation}
    \begin{aligned}
        \mathbf{J}^Q_T=T\bar{s}\mathbf{u}+\frac{e^2}{\hbar}k_B T\mathbf{E}\times\int{[d\mathbf{p}]\left[\frac{\epsilon_{\mathbf{p}}-\mu}{k_B T} f^0+\log(1+e^{-\beta(\epsilon_\mathbf{p}-\mu)})\right]}+\frac{1}{\hbar}\frac{\nabla T}{T}\times\int{[d\mathbf{p}]\Omega_{\mathbf{p}}\int{d\tilde{\mu}(\epsilon_{\mathbf{p}}-\tilde{\mu})^2\frac{\partial f^0}{\partial \tilde{\mu}}}}+\\\frac{1}{\hbar}\int{[d\mathbf{p}]}(\epsilon_{\mathbf{p}}-\mu)\nabla(\mathbf{p.u})\times\Omega_{\mathbf{p}}f^0+\epsilon_{ijl}\partial_ju_k [M^\pi_0]_{kl}
    \end{aligned}
    \end{equation}
    \begin{equation}
        [\Pi_T]_{ij}=\int{[d\mathbf{p}]p_i\frac{\partial\epsilon_{\mathbf{p}}}{p_j}f^0}+\frac{e}{\hbar}\int{[d\mathbf{p}]p_i(\mathbf{E}\times{\Omega_{\mathbf{p}}})_j f^0}-\epsilon_{jkl}(\partial_k\mu \frac{\partial [M_0^{\pi}]_{il}}{\partial\mu}+\partial_k T\frac{\partial [M^{\pi}_0]_{il}}{\partial T})
    \end{equation}
\end{subequations}

The expressions for the charge and energy magnetization densities were derived in Ref.\cite{10.21468/SciPostPhysCore.6.3.052} by imposing that the transport currents obey Einstein's relations, i.e. transport currents must only depend on the combination $(eE+\nabla\mu)$.
We shall use this approach to find the expression for $[M^{\pi}_0]_{ij}$, the $j$-th component of $\mathbf{M^{\pi}_i}$ with $\psi=0$.
\begin{equation}
[\Pi_T]_{ij}=\int{[d\mathbf{p}]p_i\left(\frac{\partial\epsilon_{\mathbf{p}}}{\partial p_j}+\frac{e}{\hbar}\epsilon_{jkl}E_k[\Omega_{\mathbf{p}}]_l\right)}-\epsilon_{jkl}\partial_k [M^{\pi}_0]_{il}
\end{equation}

Since $[M^{\pi}_0]_{ij}=[M^{\pi}_0]_{ij}(\mu,T,\mathbf{u})$, $\partial_k [M^{\pi}_0]_{ij}=\partial_k\mu \frac{\partial [M^{\pi}_0]_{ij}}{\partial\mu}+\partial_k T \frac{\partial [M^{\pi}_0]_{ij}}{\partial T}+\partial_k u_i \frac{\partial [M^{\pi}_0]_{ij}}{\partial u_i}$.

Imposing the Einstein relation on the transport stress tensor gives:
\begin{equation}
    \frac{\partial [M^{\pi}_0]_{ij}}{\partial\mu}=-\frac{1}{\hbar}\int{[d\mathbf{p}]p_i[\Omega_{\mathbf{p}}]_l\frac{1}{e^{\beta(\epsilon_{\mathbf{p}}-\mu-\mathbf{u}.\mathbf{p})}+1}}
\end{equation}
Integrating with respect to $\mu$ gives the Berry curvature contribution to the equilibrium stress magnetization tensor.
\begin{equation}\label{eq:stmagten}
    [M^{\pi}_0]_{ij}=-\frac{1}{\hbar}k_B T\int{[d\mathbf{p}]p_i[\Omega_{\mathbf{p}}]_j\log(1+e^{-\beta(\epsilon_{\mathbf{p}}-\mu-\mathbf{u}.\mathbf{p})})}
\end{equation}

\section{Einstein relations for the vector potential}
Since the vector potential $\boldsymbol{\chi}$ couples to the momentum density, we expect there to be Einstein's relations between $\boldsymbol{\chi}$ and $\mathbf{u}$.
We derive these relations by extending the framework in Ref.~\cite{PhysRevB.55.2344} to a hydrodynamic system.
In equilibrium, the system can be described by the density matrix:
\begin{equation}
    \rho=\frac{e^{-\beta(H_T-\xi N-\boldsymbol{\zeta}.\mathbf{P})}}{\mathcal{Z}}
\end{equation}
Where $\beta=1/T_0,\xi$ are the Lagrange multipliers interpreted as the inverse temperature and electrochemical potential, respectively.
Analogously, we interpret the Lagrange multiplier $\boldsymbol{\zeta}$ as the \textit{boost velocity}.

For slowly varying fields, the system can be assumed to consist of a collection of subsystems at local equilibrium, defined by $\bar{n}(\mathbf{r}),\bar{s}(\mathbf{r}),\mathbf{\bar{g}}(\mathbf{r})$.
The local statistical fields can be defined through suitable derivatives of the entropy density:
\begin{align}\label{eq:derivatives}
    1/T=\frac{\partial \bar{s}}{\partial \bar{\epsilon}}\Bigr|_{\bar{n},\bar{\mathbf{p}}} &&
    \mu=-T\frac{\partial \bar{s}}{\partial \bar{n}}\Bigr|_{\bar{\epsilon},\bar{\mathbf{p}}} &&
    u_i=-T\frac{\partial \bar{s}}{\partial \bar{g}_i}\Bigr|_{\bar{n},\bar{\mathbf{\epsilon}}}
\end{align}
The equilibrium is obtained by minimizing the free energy functional $\mathcal{F}=E_T-\xi N-\boldsymbol{\zeta}.\mathbf{P}-T S$.
\begin{equation}
    \mathcal{F}=\int{d\mathbf{r}\left[ \left(\bar{\epsilon}_{\mathbf{r}}(1+\psi(\mathbf{r}))-e\bar{n}(\mathbf{r})\phi(\mathbf{r})+\boldsymbol{\chi}(\mathbf{r}).\mathbf{\bar{g}}(\mathbf{r})\right)-\xi\bar{n}(\mathbf{r})-\boldsymbol{\zeta}.\mathbf{\bar{g}}(\mathbf{r})-T_0 \bar{s}(\bar{\epsilon}(\mathbf{r}), n(\mathbf{r}), \mathbf{\bar{g}}(\mathbf{r}))\right]}
\end{equation}
Taking the functional derivatives gives us the relations:
\begin{align}
    1/T(\mathbf{r})=(1+\psi(\mathbf{r}))/T_0 &&
    \mu(\mathbf{r})=\frac{\xi+e\phi(\mathbf{r})}{1+\psi(\mathbf{r})}&&
    \mathbf{u}(\mathbf{r})=\frac{\boldsymbol{\zeta}-\boldsymbol{\chi}(\mathbf{r})}{1+\psi(\mathbf{r})}
\end{align}
While the first two relations have been worked out in Ref.\cite{PhysRevB.55.2344}, the last equation gives the condition on the velocity field for local equilibrium.
\begin{equation}\label{eq:eqmcondition}
    \delta(\mathbf{u}/T)=-\boldsymbol{\chi}(\mathbf{r})/T_0
\end{equation}
Where $\delta$ gives the departure from the equilibrium value.
Thus, the transport currents contain the velocity field and the vector potential in the combination $\partial_i u_j+T\partial_i(\chi_j/T)$ in order to vanish when the condition Eq.\eqref{eq:eqmcondition} is met.
Thus, to linear order, or in the absence of a non-uniform $\psi(\mathbf{r})$, the transport currents are functions of $\mathbf{u}+\boldsymbol{\chi}$.
This is the Einstein relation for the velocity field and the vector potential.

We can check that our currents obey this Einstein relation by including an anomalous velocity contribution $\frac{1}{\hbar}\nabla(\mathbf{p}.\boldsymbol{\chi})\times\Omega_{\mathbf{p}}$ arising from the mechanical force due to $\boldsymbol{\chi}$.
Thus, calculating the transport currents using Eq.~\eqref{eq:eq_mag} and Eq.~\eqref{eq:magcurr}, we get:
\begin{subequations}
\begin{equation}
    \mathbf{J^N_T}=-e\bar{n}(\mathbf{u}+\boldsymbol{\chi})-\frac{e}{\hbar}\int{[d\mathbf{p}]\nabla((\mathbf{u}+\boldsymbol{\chi}).\mathbf{p})\times\Omega_{\mathbf{p}}f^0}
\end{equation}
\begin{equation}
    \mathbf{J^Q_T}=\bar{s}T(\mathbf{u}+\boldsymbol{\chi})+\frac{1}{\hbar}\int{[d\mathbf{p}]\nabla(\mathbf{p}.(\mathbf{u}+\boldsymbol{\chi}))\left[(\epsilon_{\mathbf{p}}-\mu)f^0+k_BT\log(1+e^{-\beta(\epsilon_{\mathbf{p}}-\mu)})\right]}
\end{equation}
\begin{equation}
    [\Pi_T]_{ij}=\frac{\epsilon_{jkl}}{\hbar}\int{[d\mathbf{p}]p_i\partial_k(p.(\mathbf{u}+\boldsymbol{\chi}))[\Omega_{\mathbf{p}}]_l}
\end{equation}
\end{subequations}
Thus we see that all the transport currents are a function of $\mathbf{u}+\boldsymbol{\chi}$, verifying our Einstein relations.

\section{Einstein relations with a gravitational potential}
In addition to the Einstein relation for the electric and chemical potential, we have an Einstein relation between the gravitational potential and the temperature\cite{PhysRev.135.A1505} which states that the transport currents must only depend on the combination $\nabla\psi+\frac{\nabla T}{T}$.
Considering only gradients in $\psi$($\phi=0$).
Gradients in the gravitational potential correspond to a mechanical force, thus giving rise to an anomalous velocity $\frac{1}{\hbar}(\epsilon_{\mathbf{p}}-\mu)\nabla\psi\times\Omega_{\mathbf{p}}$.
Since our formalism has introduced the stress tensor, we shall verify the gravitational Einstein relation for the transport stress tensor in Eq.~\eqref{eq:magcurr}.
Considering only gradients in temperature and gravitational potential, the transport stress tensor takes the form:
\begin{equation}
    [\Pi_T]_{ij}=\int{[d\mathbf{p}]p_i\frac{\partial\epsilon_{\mathbf{p}}}{\partial p_j}f^0}+\frac{1}{\hbar}\int{[d\mathbf{p}]p_i[\Omega_{\mathbf{p}}]_l\epsilon_{jkl}\left[(\epsilon_{\mathbf{p}}-\mu)f^0+k_BT\log(1+e^{-\beta(\epsilon_{\mathbf{p}}-\mu)})\right]\left(\partial_k\psi+\frac{\partial_k T}{T}\right)}
\end{equation}

Clearly, the Berry curvature-dependent term obeys the Einstein relation between the gravitational potential and temperature.

\section{Calculation of the vortical coefficients for a microscopic model}
We consider a two dimensional system described by the Hamiltonian 
\begin{equation} \label{eq:ham2d}
     H = vp_x\sigma_y - \tau_z p_y \left(v\sigma_x - \alpha \right) + \Delta \sigma_z,
\end{equation}
first described in Ref.\cite{PhysRevLett.115.216806}.
$\Delta$ introduces a gap in the spectrum for the two valleys labeled by $\tau_z$, $v$ is the dispersion velocity in the absence of a gap and $\alpha$ introduces a tilt in the Dirac cones. The vector ${\bf k}$ is measured about the Dirac points for each of the two valleys. 
The energy dispersion for the two valleys with $s=\pm 1$ is
\begin{equation} \label{eq:disp}
    \epsilon_s({\bf p}) = s \alpha p_y + {\rm sgn} (\mu) \sqrt{\left(v^2p^2 + \Delta^2\right)},
\end{equation}
where the chemical potential $\mu > 0 (< 0)$ corresponds to the conduction (valence) band. The Berry curvature 
\begin{equation} \label{eq:Berry}
    \Omega_s ({\bf p}) = \frac{{\rm sgn}(\mu)} {2} \frac{sv^2\Delta}{\left(v^2p^2 + \Delta^2\right)^{3/2}}.
\end{equation}
The Berry curvature $\Omega_s ({\bf p})$ is in the $\hat{z}$ direction for both valleys. Note that $\Delta \neq 0$ breaks inversion symmetry and so that $\Omega_s({\bf p}) \neq \Omega_{-s}(-{\bf p})$. 
The expressions for the coefficients $L^3$ and $L^6$ are
\begin{subequations}
    \begin{equation}\label{eq:l3}
        L^3_{i,jk}=\frac{e}{\hbar}\epsilon_{ikl}\sum_s \int{[d\mathbf{p}]p_j[\Omega_s(\mathbf{p})]_lf^0}[\epsilon_s({\mathbf{p}})]
    \end{equation}
    \begin{equation} \label{eq:l6}
    L^6_{i,jk}=-k_B T\epsilon_{ikl}\sum_s \int[d\mathbf{p}]p_j[\Omega_s(\mathbf{p})]_l\left[\frac{\epsilon_s({\mathbf{p}})-\mu}{k_B T}f^0[\epsilon_s({\mathbf{p}})]+\log(1+e^{-\beta[\epsilon_s({\mathbf{p}})-\mu)]}\vphantom{\int_1^2}\right],
  \end{equation}
\end{subequations}
with the reciprocal relations 
\begin{subequations}\label{eq:Onsager}
    \begin{equation}
        L^3_{k,ij}=-L^7_{ij,k}
    \end{equation}
    \begin{equation}
        L^8_{ij,k}=-\frac{1}{T}L^6_{k,ij}
    \end{equation}
\end{subequations}
relating $L^7$ and $L^8$ to $L^3$ and $L^6$.
The integrals in Eqns.~\ref{eq:l3} and~\ref{eq:l6} are of the sums of the integrals for each of the two valleys labeled by $s=\pm 1$. 

From the symmetry of the problem, the only non-zero coefficients are $L^3_{x,yy}$ and $L^6_{x,yy}$. An analytical expression for the coefficient $L^3_{x,yy}$ can be obtained in the limit $\alpha \ll (v, \Delta/p_F)$ at $T=0$ and is given below 

\begin{equation} \label{eq:vort1}
L^3_{x,yy} = -\frac{e}{4\pi\hbar} \frac{ \Delta \alpha (\mu^2-\Delta^2)}{v^2\mu^2},
\end{equation}
It can be seen that $L^3_{x,yy} \rightarrow 0$ as $\alpha \rightarrow 0$ showing that while breaking inversion symmetry is a necessary condition for $L^3$ to be non-zero, it is not sufficient (inversion symmetry for the system we have considered, is broken even for $\alpha=0$). $L^3$ for other values of the parameters can be evaluated by performing the integral in Eqn.~\ref{eq:l3} numerically and the result of such a calculation is shown in the main text. 

For small $\alpha, k_B T$, we can treat the tilt as a perturbation to an isotropic spectrum to get the leading-order contribution to $L^6_{xyy}$, which is given by:
\begin{equation}
    L^6_{xyy}=\frac{k_B T}{\hbar}\int{[d\mathbf{p}]p_y\frac{sv^2\Delta}{2(v^2p^2+\Delta^2)^{3/2}}\left[\frac{\epsilon_{\mathbf{p}}-\mu}{k_B T}f^0+\log(1+e^{-\beta(\epsilon_{\mathbf{p}}-\mu)})\right]}
\end{equation}

\begin{figure*}
\includegraphics[width=\textwidth]{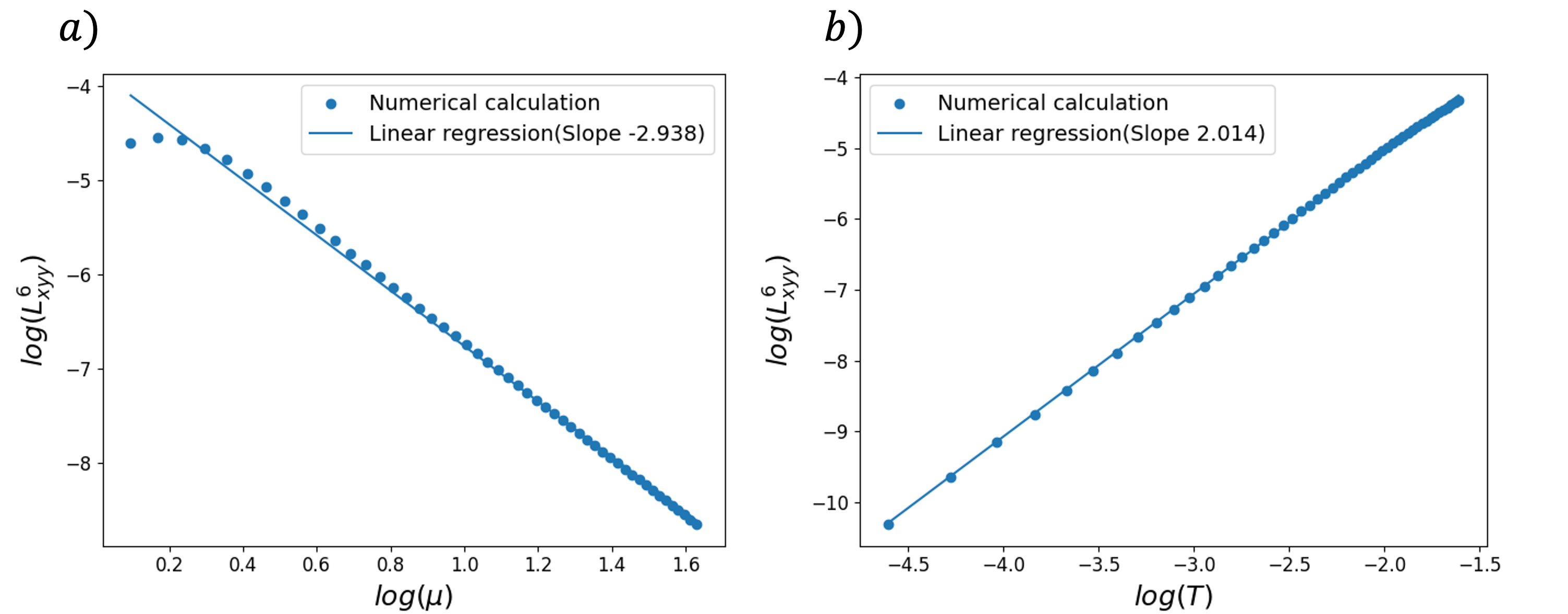}    
\centering
\caption{Extracting the asymptotic $\mu$(a) and $T$(b) dependence of $L^6_{xyy}$ from the log-log profiles. The scatter plot shows the numerically calculated values, and the linear plot shows the linear regression with the slope given in the inset.
The numerical calculations were performed with $\alpha=0.1$ and $\Delta=1eV$. Since the expression Eq.~\eqref{eq:l6approx} is only valid for $\alpha\ll\mu$, we see that the agreement with the inverse-cubed power law improves for higher $\mu$ in (a).
Similarly, the agreement with the squared power-law for $T$ becomes worse in (b) at higher $T$ since we need to include higher-order terms in the Sommerfeld expansion.}
\label{fig:curvefits}
\end{figure*}

Expanding about the isotropic dispersion $\epsilon_{\mathbf{p}}=\epsilon_{\mathbf{p}}^0+s\alpha p_y$, where $\epsilon_{\mathbf{p}}^0=\sqrt{v^2p^2+\Delta^2}$, and summing over $s=\pm 1$.
\begin{equation}
    L^6_{xyy}= \frac{v^2\Delta}{\hbar} \int{[d\mathbf{p}]\frac{p_y^2 \alpha v^2\Delta}{(v^2p^2+\Delta^2)^{3/2}}(\epsilon_{\mathbf{p}}-\mu)\left(-\frac{\partial f^0}{\partial \epsilon_{\mathbf{p}^0}}\right)}
\end{equation}
Using the isotropy of $\epsilon^0_{\mathbf{p}}$, we can replace the $p_y^2$ with $p^2/2$ and then use the Sommerfeld expansion to get the leading order contribution:
\begin{equation}\label{eq:l6approx}
    L^6_{xyy}=\frac{\pi\alpha \Delta^3}{6\hbar\mu^3v^2}(k_B T)^2
\end{equation}

We numerically recover this behaviour for $L^6_{xyy}$ as a function of the different parameters to get $L^6_{xyy}\approx \alpha T^{2.014}/\mu^{2.938}$, shown in Fig.~\ref{fig:curvefits} which are very close to the exponents expected.
The result of such a calculation is also shown in the main text.

\bibliography{references}